\documentclass[12pt]{iopart}

\usepackage{graphicx}
\begin{document}

\title[Phase transitions of an intrinsic curvature model]{Phase transitions of an intrinsic curvature model on dynamically triangulated spherical surfaces with point boundaries}

\author{S. Obata, M. Egashira, T. Endo, and  H. Koibuchi}

\address{Department of Mechanical and Systems Engineering, Ibaraki National College of Technology, Nakane 866 Hitachinaka, Ibaraki 312-8508, Japan}
\ead{koibuchi@mech.ibaraki-ct.ac.jp}

\begin{abstract}
An intrinsic curvature model is investigated using the canonical Monte Carlo simulations on dynamically triangulated spherical surfaces of size upto $N\!=\!4842$ with two fixed-vertices separated by the distance $2L$. We found a first-order transition at finite curvature coefficient $\alpha$, and moreover that the order of the transition remains unchanged even when $L$ is enlarged such that the surfaces become sufficiently oblong. This is in sharp contrast to the known results of the same model on tethered surfaces, where the transition weakens to a second-order one as $L$ is increased. The phase transition of the model in this paper separates the smooth phase from the crumpled phase. The surfaces become string-like between two point-boundaries in the crumpled phase. On the contrary, we can see a spherical lump on the oblong surfaces in the smooth phase. The string tension was calculated and was found to have a jump at the transition point. The value of $\sigma$ is independent of $L$ in the smooth phase, while it increases with increasing $L$ in the crumpled phase. This behavior of $\sigma$ is consistent with the observed scaling relation $\sigma \sim (2L/N)^\nu$, where $\nu\simeq 0$ in the smooth phase, and $\nu\!=\!0.93\pm 0.14 $ in the crumpled phase. We should note that a possibility of a continuous transition is not completely eliminated. 

\end{abstract}

\maketitle

\section{Introduction}\label{intro}
Phase transition of surfaces has been investigated to understand string models and biological membranes \cite{WHEATER-JP1994,NELSON-SMMS2004,David-TDQGRS-1989,David-SMMS2004,Wiese-PTCP2000,Bowick-PREP2001,Gompper-Schick-PTC-1994}. Efforts has been devoted to clarify the phase structure \cite{Peliti-Leibler-PRL1985,DavidGuitter-EPL1988,PKN-PRL1988,BKS-PLA2000,BK-PRB2001}  in the model of Helfrich, Polyakov, and Kleinert (HPK), which is defined by a two-dimensional curvature Hamiltonian \cite{HELFRICH-1973,POLYAKOV-NPB1986,KLEINERT-PLB1986}. The self-avoiding property of surfaces are crucial for biological membranes \cite{GREST-JPIF1991,BOWICK-TRAVESSET-EPJE2001,BCTT-PRL2001}, however, it is less important for string models \cite{WHEATER-JP1994}. A surface model that is allowed to self-intersect is conventionally called a phantom surface model, which has long been studied numerically on tethered surfaces \cite{KANTOR-NELSON-PRA1987,KANTOR-SMMS2004,WHEATER-NPB1996} and on fluid surfaces \cite{BCFTA-JP96-NPB9697,CATTERALL-NPBSUP1991,AMBJORN-NPB1993,ABGFHHM-PLB1993,BCHHM-NPB9393,KOIB-PLA20023,KOIB-PLA-2004,KOIB-EPJB-2005,KOIB-EPJB-2006}. Recent numerical simulations on phantom HPK models revealed that there is a first-order transition between the smooth phase and the crumpled phase \cite{KD-PRE2002,KOIB-PRE-2004-1,KOIB-PRE-2005-1,KOIB-NPB-2006}. The results in \cite{KOIB-PRE-2005-1,KOIB-NPB-2006} imply a possible phase transition in biological or artificial membranes, because the Hausdorff dimension $H$ is $H<3$ in the crumpled phase even though the surface is allowed to self-intersect.

The string tension $\sigma$ is a key observable to understand the phase structure of surfaces \cite{AMBJORN-NPB1993}. If a surface becomes oblong, string-like, and of length $2L$, the free energy of the surface can be expressed by $\sigma 2L$. Therefore, $\sigma$ can be extracted from the grand canonical ensembles of surfaces with fixed boundaries of sufficiently large distance $2L$. It was reported that an extrinsic curvature model undergoes a first-order transition, where the string tension becomes non-zero in the smooth phase close to the transition point in the limit of $L \to \infty$ \cite{KOIB-PLA-2004,KOIB-EPJB-2005}. The results of the extrinsic curvature model in \cite{KOIB-PLA-2004,KOIB-EPJB-2005} show that the order of the transition changes depending on $L$.

However, it is still unknown how the distance $2L$ between the boundaries influences the phase transition of an intrinsic curvature model on dynamically triangulated surfaces. The intrinsic curvature model was first studied by Baillie et.al. \cite{BJ-PRD-1993-1994,BEJ-PLB-1993,BIJJ-PLB-1994,FW-PLB-1993}. A surface model with deficit angle term, which is an intrinsic curvature model, has a first-order transition on tethered surfaces \cite{KOIB-PRE-2004-2}. Another intrinsic curvature model undergoes a first-order transition, which is independent of the surface topology \cite{KOIB-EPJB-2004,KOIB-PLA-2005-1,KOIB-PLA-2005-2}. It was also reported that the order of transition changes from first to second as $L$ increases from $L\!=\!L_0$ to $L\!=\!2L_0$, where $L_0$ is the radius of the original sphere constructed to satisfy the relation $S_1/N\!=\!1.5$ \cite{KOIB-JSTAT-2006-1}. However, the string tension was unable to be extracted from the simulations on the tethered surfaces, because the surfaces do not always become string-like even when $L$ is relatively large compared to $L_0$. On the contrary, the fluid surface can be oblong in the limit of $L\to\infty$. Therefore, we expect that not only the dependence of the transition on the distance $2L$ but also the string tension $\sigma$ is evaluated from the intrinsic curvature models on fluid surfaces with the fixed boundary-vertices.

In this paper, we will study the dependence of the transition on the distance $2L$ of the boundary of spherical fluid surfaces by MC simulations with dynamical triangulations. The string tension can be obtained by the canonical MC technique, although it is originally defined by the grand canonical ensemble. We confine ourselves into a phantom surface model in this paper. However, it is natural to expect the surfaces are not completely self-intersected even in the crumpled phase, because they span a long distance between two point-boundaries and consequently become stretched out.

\section{Model}
The Gaussian bond potential $S_1$ and the intrinsic curvature energy $S_3$ are defined by
\begin{equation}
\label{Disc-Eneg} 
S_1=\sum_{(ij)} \left(X_i-X_j\right)^2,\quad S_3=-\sum_i\log( \delta_i/2\pi),
\end{equation} 
where $\sum_{(ij)}$ in $S_1$ is the sum over bonds $(ij)$ connecting the vertices $i$ and $j$. The bonds $(ij)$ are the edges of the triangles. $\delta_i$ in Eq. (\ref{Disc-Eneg}) is the vertex angle, which is the sum of the angles meeting at the vertex $i$. We call $S_3$ the deficit angle term, because $\delta_i\!-\!2\pi$ is just the deficit angle. We note that $\sum_i (\delta_i\!-\!2\pi)$ is constant on surfaces of fixed genus because of the Gauss-Bonnet theorem. 

The partition function of the fluid surface model is defined by 
\begin{eqnarray} 
\label{Part-Func}
 Z = \sum_{\cal T}\int^\prime \prod _{i=1}^{N} d X_i \exp\left[-S(X,{\cal T})\right],\\  
 S(X,{\cal T})=S_1 + \alpha S_3, \nonumber
\end{eqnarray} 
where $\alpha$ is the curvature coefficient, and $\sum_{\cal T}$ denotes the summation over all possible triangulation ${\cal T}$. $\int^\prime$ in Eq.(\ref{Part-Func}) denotes the boundary condition in which two vertices are fixed and separated by the distance $2L(N)$, which will be discussed in the following section in more detail. $S(X,{\cal T})$ denotes that the Hamiltonian $S$ depends on the position variables $X$ of vertices and the triangulation ${\cal T}$.

We should comment on the unit of physical quantities in the model. We can express all physical quantities by unit of length in terms of $a$,  which is a length unit in the model. Hence, the unit of $S_1$ and $L_0(N)$ are $a^2$ and $a$, respectively. We fix the value of $a$ to $a\!=\!1$ in this paper, because the length unit can be arbitrarily chosen on the basis of the scale invariant property in the model. The unit of $\alpha$ is expressed by $kT$, where $k$ is the Boltzmann constant, $T$ is the temperature.  Note also that varying the temperature $T$ is effectively identical with varying the curvature coefficient $\alpha$ in the model. 

We note also a relation between the deficit angle term $S_3$ and the integration measure $\Pi_i d X_i q_i^{3/2}$ in the partition function of a fluid surface model \cite{DAVID-NPB-1985}, where $q_i$ is the coordination number of the vertex $i$. The weight factor $q_i^{3/2}$ can also be written as $\exp [(3/2)\sum_i\log q_i]$. Then, replacing $q_i$ by $\delta_i$ and $3/2$ by $\alpha$ the continuous number, we have $\exp [\alpha \sum_i\log \delta_i]$. Introducing the constant term into the Hamiltonian, we have $S_3$ in Eq.(\ref{Disc-Eneg}).       

\section{Triangulated spheres and Monte Carlo technique} \label{MC-technique}
Triangulated surfaces are obtained by dividing the icosahedron. By splitting the edges of the icosahedron into $\ell$-pieces, we have a surface of size $N\!=\!10\ell^2\!+\!2$. These surfaces are characterized by $N_5\!=\!12$ and $N_6\!=\!N\!-\!12$, where $N_q$ is the total number of vertices with a co-ordination number $q$. 

The spherical surfaces are constructed by fixing the radius $L_0(N)$ so that the Gaussian potential $S_1/N$ is approximately equal to $S_1/N\!\simeq\!3/2$. We assume the following values of $L_0(N)$: $L_0(N)\!=\!7$ for $N\!=\!1442$, $L_0(N)\!=\!9.3$ for $N\!=\!2562$, $L_0(N)\!=\!12.8$ for $N\!=\!4842$. Under those values of $L_0(N)$, the relation $S_1/N\!=\!1.5$ is almost satisfied on those spheres. Note that the value of the radius $L_0(N)$, which satisfies $S_1/N\!\simeq\!3/2$, may slightly change depending on how to construct the sphere. 

The distance $2L(N)$ between the boundary vertices is fixed to three different values such that
\begin{equation}
\label{Distance} 
L(N)=2L_0(N), \; 3L_0(N), \; 5L_0(N)  
\end{equation} 
where $L_0(N)$ is the radius of sphere constructed from the icosahedron as described above.

\begin{figure}[hbt]
\vspace{1cm}
\unitlength 0.1in
\begin{picture}( 0,0)(  20,28)
\put(27,52.5){\makebox(0,0){(a) $L(N)\!=\!2L_0(N)$}}%
\put(26,50.5){\makebox(0,0){$N\!=\!4842$}}%
\put(47,52.5){\makebox(0,0){(b) $L(N)\!=\!3L_0(N)$}}%
\put(46,50.5){\makebox(0,0){$N\!=\!4842$}}%
\put(67,52.5){\makebox(0,0){(c) $L(N)\!=\!5L_0(N)$}}%
\put(66,50.5){\makebox(0,0){$N\!=\!4842$}}%
\end{picture}%
\centering
\includegraphics[width=15.5cm]{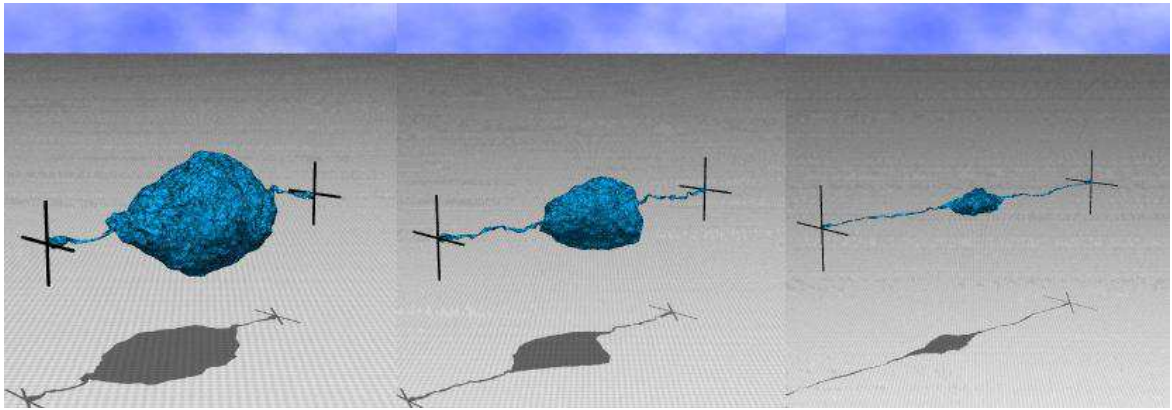}
\caption{Starting configurations under the conditions (a) $L(N)\!=\!2L_0(N)$, (b) $L(N)\!=\!3L_0(N)$, and (c)  $L(N)\!=\!5L_0(N)$. The symbols $+$ at the terminal points represent the position of the boundary vertices. The distance between the boundaries is $2L(N)$ in each surface. These figures are drawn in scales which are different from each other. }
\label{fig-1}
\end{figure}

Figures \ref{fig-1}(a), \ref{fig-1}(b), and \ref{fig-1}(c) show the starting configurations of the MC simulations for $L(N)\!=\!2L_0(N)$, $L(N)\!=\!3L_0(N)$, and $L(N)\!=\!5L_0(N)$. In order to show the position of boundary points, the symbols $+$ are plotted at the points in each figure. These figures are drawn in scales which are different from each other. The size of spherical lump at the center of each snapshot is a little bit smaller than that of the original sphere of radius $L_0(N)$. 

The surfaces in Figs.\ref{fig-1}(a)--\ref{fig-1}(c) are obtained by pulling the boundary vertices out of the original spheres. We choose the boundary points and fix the positions as follows: Firstly, we choose a pair of vertices, which are on a straight line passing through the center of the sphere. The canonical $x$-coordinate axis in ${\bf R}^3\!=\!\{(x,y,z)\vert x,y,z \in {\bf R}\}$ is chosen along the straight line. The distance between these two vertices on the axis is given by $2L_0(N)$ at the beginning. Secondly, the distance is enlarged from $2L_0(N)$ to $2L(N)$ along $x$-axis during the thermalization MCS (Monte Carlo Sweep), where $L(N)$ is given in Eq. (\ref{Distance}). The vertices are pulled away from the sphere during the first $0.25\!\times\!10^7$ thermalization MCS,  and then $0.75\!\times\!10^7$ or more thermalization MCS is performed after the expansion was finished.  Equilibrium statistical mechanical conditions seem to be violated slightly in the first $0.25\!\times\!10^7$ MCS because of this forced expansion of vertices. However, it should be noted that these MCS are iterated only for the constructions of starting configurations. These MCS are neither the production MCS for sampling nor the thermalization MCS. Therefore, the final results are independent of the method to get the starting configurations because the number of thermalization MCS is large enough. 

The distance $2L(N)$ is obtained by enlarging the distance between the boundary points from $2L(N)\!=\!14$ to $2L(N)\!=\!28$ on the $N\!=\!1442$ surface for example. Two fixed vertices are shifted by distance $7\times 10^{-5}$ at every 25 MCS to the opposite direction to each other along $x$-axis during the first $0.25\times 10^7$ MCS. We must emphasize that the results of the simulation is completely independent of how the surface with the boundaries of distance $2L(N)$ is constructed from the sphere of diameter $2L_0(N)$. 

The distance between boundary vertices on the surface of $L(N)\!=\!5L_0(N)$ is just $10L_0(N)$ for example. The boundary conditions are thus imposed on the model by these two fixed vertices of distance $2L(N)$. If it were not for the boundary condition, we have $S_1/N\!=\!3/2$, which comes from the scale invariant property of $Z$. However, as we will see later, this relation is slightly broken due to the boundary condition.

The vertices $X$ are shifted so that $X^\prime \!=\! X\!+\!\delta X$ in MC, where $\delta X$ is chosen randomly in a small sphere. The new position $X^\prime$ is accepted with the probability ${\rm Min}[1,\exp(-\Delta S)]$, where $\Delta S\!=\! S({\rm new})\!-\!S({\rm old})$. We use a sequence of random numbers called Mersenne Twister \cite{Matsumoto-Nishimura-1998} for three dimensional random shift of $X$ and for the Metropolis accept/reject. The radius of the small sphere for the shift $\delta X$ is chosen so that the rate of acceptance for $X$ is about $50\%$. We introduce the lower bound $1\times 10^8$ for the area of triangles. No lower bound is imposed on the bond length. 

The dynamical triangulation is performed by the standard bond flip technique for integrating the degrees of freedom ${\cal T}$ in the partition function $Z$ of Eq.(\ref{Part-Func}). The bonds are labeled with sequential numbers, where the total number of bonds $N_B$ is given by $N_B\!=\!3N\!-\!6$. The odd-numbered bonds are firstly chosen to be flipped, and secondly the remaining even-numbered bonds are chosen. The flip is accepted with the probability ${\rm Min} [1, \exp(-\Delta S)]$. $N$-updates for $X$ and $N_B/2$-updates for ${\cal T}$ are consecutively performed and make one MCS.

In order to improve computational efficiency, we renumber the vertices in order close to one boundary vertex at every $0.5\times 10^5\sim 5\times 10^5$ MCS. Vertices freely diffuse over the surface because of the fluidity introduced by the bond flip in dynamically triangulated MC simulations. As a consequence, the vertex number is eventually distributed at random over the surface. The computations are very time consuming on such random surfaces.
This is typical of the computations on dynamically triangulated fluid surfaces. However, the renumbering the vertices can save the computer time. The speed of computation is increased about $100\%$ or more by the renumbering on the $N\!=\!4842$ surface.

\section{Results}
Monte Carlo simulations were performed on the surfaces of size $N\!=\!1442$,  $N\!=\!2562$, and $N\!=\!4842$. The distance between the boundaries is $2L(N)\!=\!4L_0(N)$, $2L(N)\!=\!6L_0(N)$, and $2L(N)\!=\!10L_0(N)$ on each surface. The total number of MCS is about $0.8\times 10^8\sim 1.5\times 10^8$ at transition point $\alpha_c$ after the thermalization MCS. Relatively small numbers of sweeps were performed at non-transition points $\alpha\not\!=\! \alpha_c $.

\begin{figure}[hbt]
\vspace{1cm}
\unitlength 0.1in
\begin{picture}( 0,0)(  20,28)
\put(30,52.5){\makebox(0,0){(a) $\alpha\!=\!1850,\;L(N)\!=\!2L_0(N)$}}%
\put(53,52.5){\makebox(0,0){(b) $\alpha\!=\!1900,\;L(N)\!=\!2L_0(N)$}}%
\end{picture}%
\centering
\includegraphics[width=11.5cm]{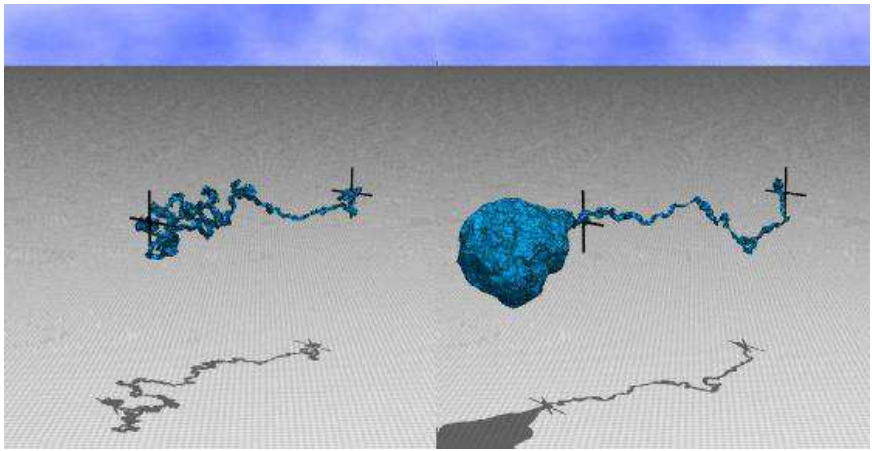}
\caption{Snapshots of the $L(N)\!=\!2L_0(N)$ surface obtained at (a) $\alpha\!=\!1850$ (crumpled phase) and (b) $\alpha\!=\!1900$ (smooth phase). The symbols $+$ denote the position of the boundary points. The snapshots are drawn in the same scale. The surface size is $N\!=\!4842$.   }
\label{fig-2}
\end{figure}
Firstly, we show snapshots of the $L(N)\!=\!2L_0(N)$ surfaces obtained at $\alpha\!=\!1850$ and $\alpha\!=\!1900$ in Figs.\ref{fig-2}(a) and  \ref{fig-2}(b), respectively. The distance between the boundaries is $2L(N)\!=\!4L_0(N)$. The surface size is  $N\!=\!4842$ in both snapshots. The surface in Fig.\ref{fig-2}(a) is crumpled and is typical of the crumpled phase, while the surface in Fig.\ref{fig-2}(b) is partly swollen and is typical of the smooth phase. The crumpled surface seems linear rather than branched-polymer, while the smooth surface has a lump of smooth sphere, which stays outside the boundary points and touches one boundary point. Almost all vertices are included in the lump. We confirmed that the lump is a smooth surface by slicing it into small sections.

\begin{figure}[hbt]
\vspace{1cm}
\unitlength 0.1in
\begin{picture}( 0,0)(  20,28)
\put(30,52.5){\makebox(0,0){(a) $\alpha\!=\!1900,\;L(N)\!=\!3L_0(N)$}}%
\put(53,52.5){\makebox(0,0){(b) $\alpha\!=\!1950,\;L(N)\!=\!3L_0(N)$}}%
\end{picture}%
\centering
\includegraphics[width=11.5cm]{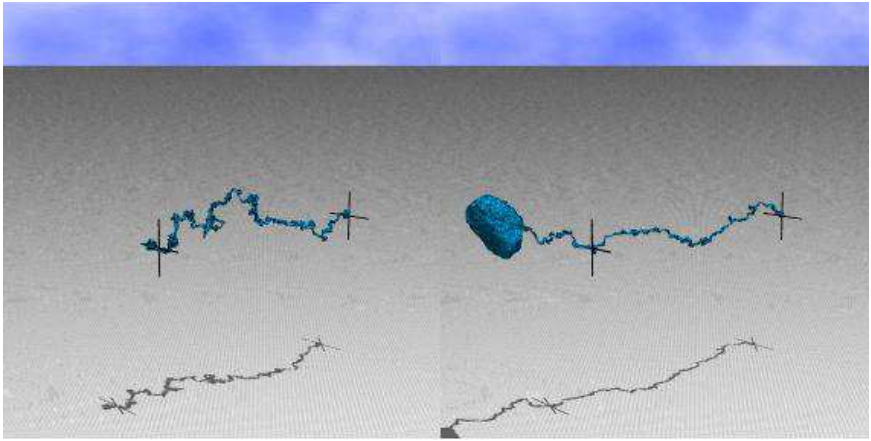}
\caption{Snapshots of the $L(N)\!=\!3L_0(N)$ surface obtained at (a) $\alpha\!=\!1900$ (crumpled phase) and (b) $\alpha\!=\!1950$ (smooth phase). The symbols $+$ denote the position of the boundary points. The snapshots are drawn in the same scale. The surface size is $N\!=\!4842$. }
\label{fig-3}
\end{figure}
Figures \ref{fig-3}(a) and \ref{fig-3}(b) are the snapshots obtained at $\alpha\!=\!1900$ and $\alpha\!=\!1950$ with the distance $2L(N)\!=\!6L_0(N)$, where $N\!=\!4842$. Both figures are drawn in the same scale. The snapshot in  Fig.\ref{fig-3}(a) is the one in the crumpled phase, while that of Fig.\ref{fig-3}(b) is the one in the smooth phase. We can also see that the crumpled surface is almost linear and the smooth surface includes a smooth lump as in Fig. \ref{fig-2}(b) for $L(N)\!=\!2L_0(N)$. The lump in the smooth surface is located outside the boundary points just as in Fig. \ref{fig-2}(b) for $L(N)\!=\!2L_0(N)$. 

\begin{figure}[hbt]
\vspace{1cm}
\unitlength 0.1in
\begin{picture}( 0,0)(  20,28)
\put(30,52.5){\makebox(0,0){(a) $\alpha\!=\!2000,\;L(N)\!=\!5L_0(N)$}}%
\put(53,52.5){\makebox(0,0){(b) $\alpha\!=\!2050,\;L(N)\!=\!5L_0(N)$}}%
\end{picture}%
\centering
\includegraphics[width=11.5cm]{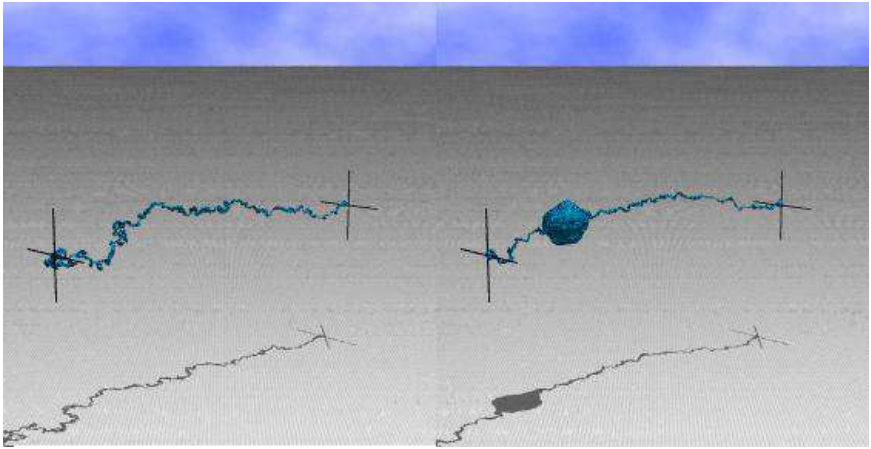}
\caption{Snapshots of the $L(N)\!=\!5L_0(N)$ surface obtained at (a) $\alpha\!=\!2000$ (crumpled phase) and (b) $\alpha\!=\!2050$ (smooth phase). The symbols $+$ denote the position of the boundary points. The snapshots are drawn in the same scale. The surface size is $N\!=\!4842$. }
\label{fig-4}
\end{figure}
Snapshots obtained at $\alpha\!=\!2000$  and  $\alpha\!=\!2050$ are shown in Figs.\ref{fig-4}(a) and \ref{fig-4}(b), where the distance between boundaries is  $2L(N)\!=\!10L_0(N)$, and the surface size is $N\!=\!4842$. The crumpled surface in Fig.\ref{fig-4}(a) is apparently linear, and the smooth surface in Fig.\ref{fig-4}(b) includes a spherical lump as those in the cases $L(N)\!=\!2L_0(N)$ and $L(N)\!=\!3L_0(N)$ in Fig.\ref{fig-2}(b) and  Fig.\ref{fig-3}(b). Contrary to the previous cases in Figs.\ref{fig-2}(b) and \ref{fig-3}(b), the spherical lump stays between the fixed boundaries in Fig.\ref{fig-4}(b). However, the lump is not yet trapped at either boundary. Surfaces span inside the fixed boundaries in both crumpled and smooth phases in the case $L(N)\!=\!5L_0(N)$.

\begin{figure}[hbt]
\centering
\includegraphics[width=12.5cm]{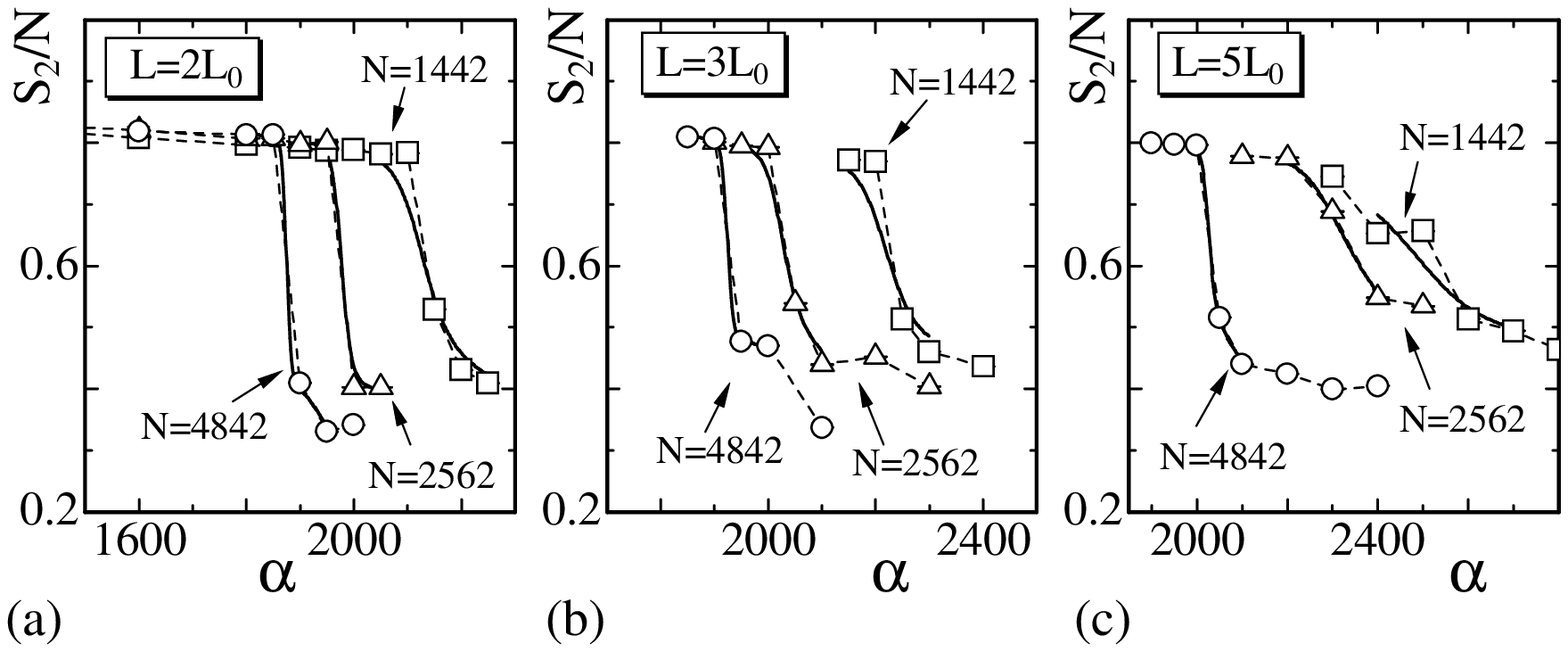}
\caption{The bending energy $S_2/N_B$ vs. $\alpha$ obtained under the conditions (a) $L(N)\!=\!2L_0(N)$, (b) $L(N)\!=\!3L_0(N)$, and (c) $L(N)\!=\!5L_0(N)$. $N_B$ is the total number of bonds. Solid lines were drawn by the multihistogram reweighting technique. }
\label{fig-5}
\end{figure}
The bending energy $S_2$ is defined by using the unit normal vector ${\bf n}_i$ of the triangle $i$ such that 
\begin{equation}
\label{S2}
S_2=\sum_{ij} (1-{\bf n}_i \cdot {\bf n}_j),
\end{equation}
where $\sum_{ij}$ denotes the sum over all nearest neighbor triangles $i$ and $j$ that share the common bond. $S_2$ is not included in the Hamiltonian, however, it reflects how smooth the surface is.

Figures \ref{fig-5}(a)--\ref{fig-5}(c) show $S_2/N_B$ against $\alpha$ obtained on the surfaces of  $L(N)\!=\!2L_0(N)$,  $L(N)\!=\!3L_0(N)$, and  $L(N)\!=\!5L_0(N)$.
 $N_B$ is the total number of bonds. Solid lines were drawn by the multihistogram reweighting technique \cite{Janke-histogram-2002}.  The reweighting is done by using $4\sim5$ data close to the transition point. The dashed lines, which simply connect the data, are drawn to guide the eyes. We can obviously see that $S_2/N_B$ discontinuously changes against $\alpha$ in all three conditions for $L(N)$. A phase transition can be called a first-order one if some physical quantity discontinuously changes. Therefore, the discontinuous change of $S_2/N_B$ in Fig.\ref{fig-5}(a)--\ref{fig-5}(c) indicates that the model undergoes a first-order transition under the conditions $L(N)\!=\!2L_0(N)$, $L(N)\!=\!3L_0(N)$, and  $L(N)\!=\!5L_0(N)$. 

The standard deviations in $S_2/N_B$ are calculated and plotted in the figures as error bars. However, the errors are small and hardly to be seen. In the following, the standard deviations are also calculated and plotted as error bars in all the figures just as those in Fig.\ref{fig-5}.

\begin{figure}[hbt]
\centering
\includegraphics[width=12.5cm]{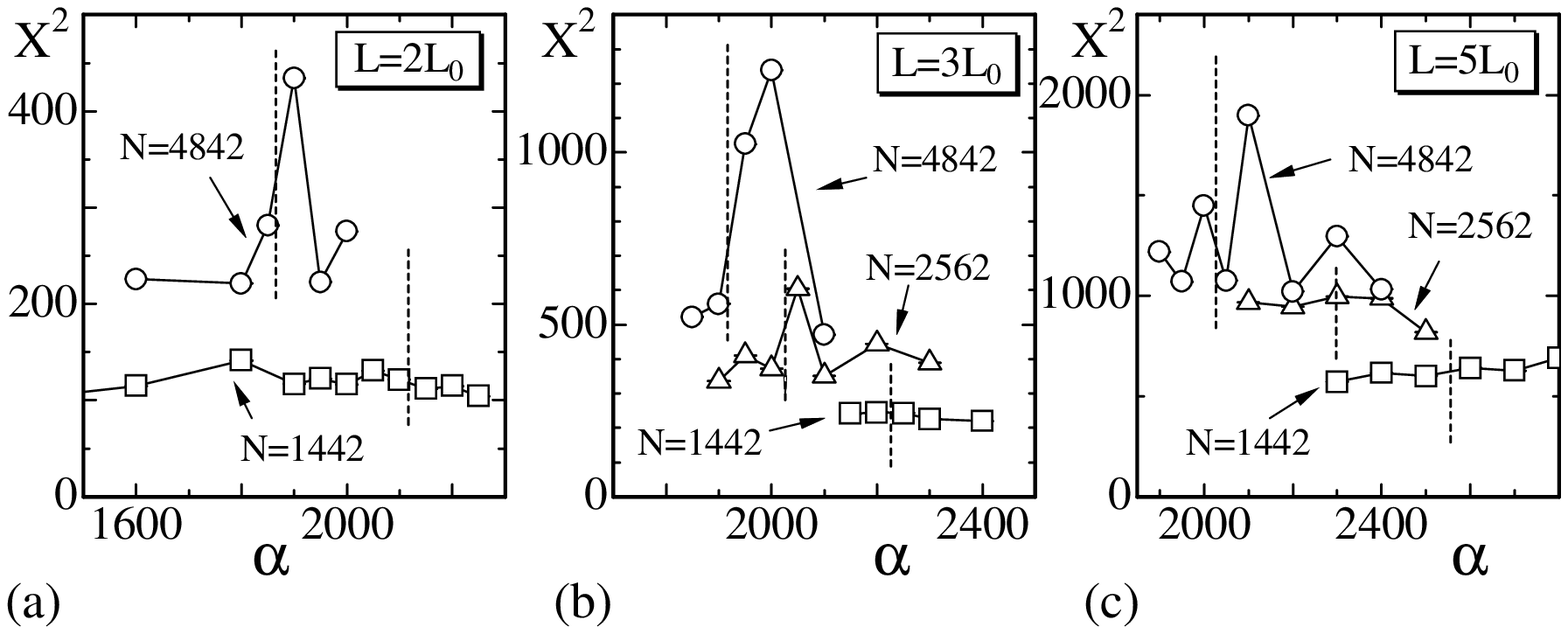}
\caption{The mean square size $X^2$ vs. $\alpha$ obtained under the conditions (a) $L(N)\!=\!2L_0(N)$ , (b) $L(N)\!=\!3L_0(N)$ , and (c) $L(N)\!=\!5L_0(N)$. Dashed lines drawn vertically denote the phase boundary between the smooth phase and the crumpled phase. }
\label{fig-6}
\end{figure}
The mean square size $X^2$ is defined by 
\begin{equation}
\label{X2}
X^2={1\over N} \sum_i \left(X_i-\bar X\right)^2, \quad \bar X={1\over N} \sum_i X_i,
\end{equation}
where $\bar X$ is the center of the surface. $X^2$ reflects the size of surfaces even on surfaces with fixed boundaries of distance $L(N)$ given by Eq. (\ref{Distance}). In the crumpled phase, $X^2$ is expected to be almost constant against $\alpha$, because the surface spans from one boundary point to the other and is almost linear, and therefore there is no reason for large fluctuations of $X^2$. On the contrary, $X^2$ largely fluctuates in the smooth phase because it depends on the position of the spherical lump, which can be seen in the snapshots shown in Figs. \ref{fig-2}(b), \ref{fig-3}(b), and \ref{fig-4}(b). $X^2$ is expected to remain small when the lump stays in the middle of the two boundary points, while it becomes relatively large if the lump moves apart from the center of the boundary points.

Figures \ref{fig-6}(a)--\ref{fig-6}(c) show  $X^2$ against $\alpha$ for $L(N)\!=\!2L_0(N)$,  $L(N)\!=\!3L_0(N)$,  and $L(N)\!=\!5L_0(N)$. The dashed lines drawn vertically in the figures denote the phase boundary between the crumpled phase and the smooth phase. We find from the figures that an expected large fluctuation can be seen in $X^2$ of the $N\!=\!4842$ surfaces at the smooth phase. Whereas $X^2$ of the $N\!=\!1442$ surfaces remains almost constant against $\alpha$. The reason of this is the size effect. The total number of vertices in the spherical lump, which appear on the surface in the smooth phase, is relatively small, and therefore $X^2$ is not influenced by where the lump is in such small sized surfaces.

\begin{figure}[hbt]
\centering
\includegraphics[width=12.5cm]{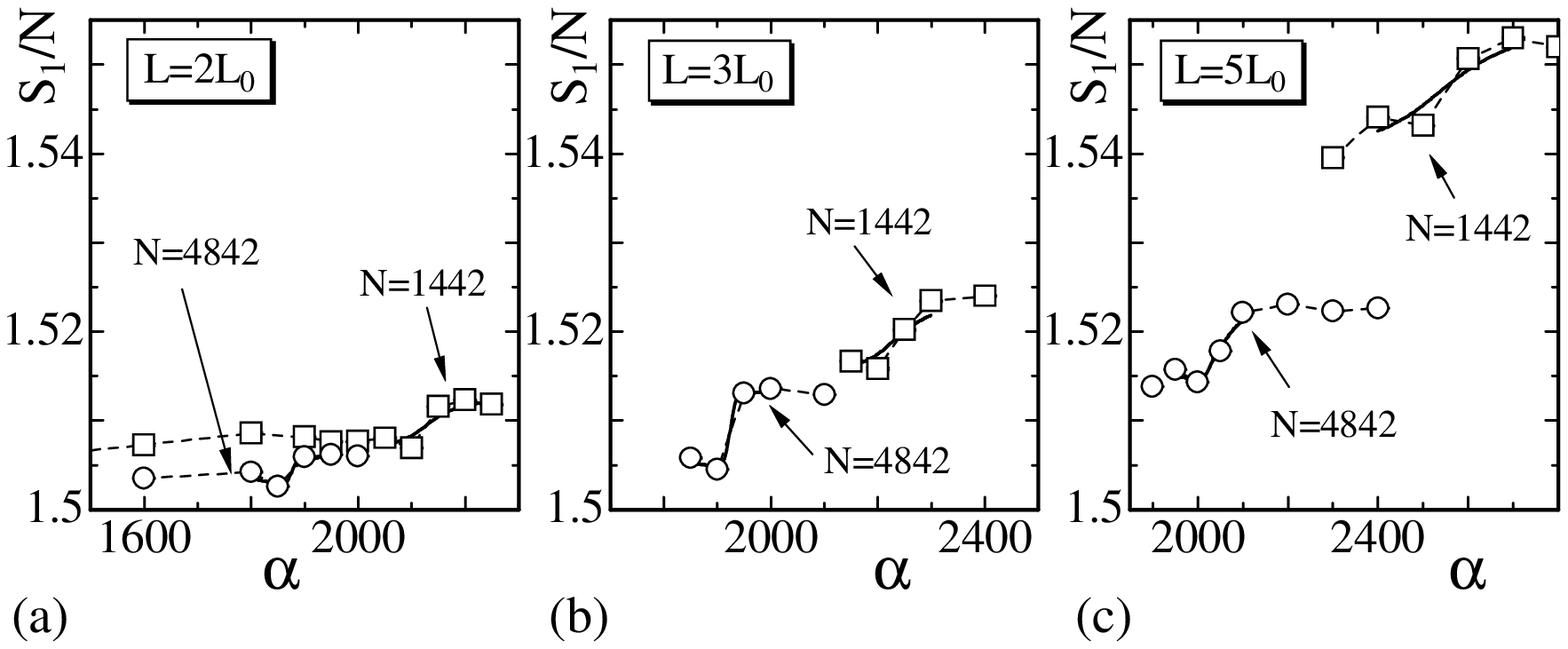}
\caption{The Gaussian bond potential $S_1$ vs. $\alpha$ obtained under the conditions (a) $L(N)\!=\!2L_0(N)$ , (b) $L(N)\!=\!3L_0(N)$ , and (c) $L(N)\!=\!5L_0(N)$. Solid lines were drawn by the multihistogram reweighting technique. }
\label{fig-7}
\end{figure}
We plot the Gaussian bond potential $S_1/N$ against $\alpha$ in Figs.\ref{fig-7}(a)--\ref{fig-7}(c). $S_1/N$ should be equal to $3/2$ whenever no specific boundary condition is imposed on the model as mentioned previously. However, $S_1/N$ in Figs.\ref{fig-7}(a)--\ref{fig-7}(c) obviously deviates from $3/2$. The reason of this is because the surface is fixed with two vertices separated by the distance $2L(N)$. The figures were plotted in the same scale along the vertical axis. We find from the figures that the deviation of $S_1/N$ from $3/2$ increases with increasing $L(N)$. It is also clear that $S_1/N$ discontinuously changes in Figs.\ref{fig-7}(a)--\ref{fig-7}(c). The discontinuity of $S_1/N$ indicates that the model undergoes a first-order transition under $L(N)\!=\!2L_0(N)$, $L(N)\!=\!3L_0(N)$, and  $L(N)\!=\!5L_0(N)$, in accordance with the prediction from the behavior of $S_2/N_B$ in Figs.\ref{fig-5}(a)--\ref{fig-5}(c).

The shape of surfaces is linear and seems string-like when the surfaces are in the crumpled phase as we have seen in the snapshots. Moreover, the surfaces are considered to be sufficiently oblong even when they are in the smooth phase except under $L(N)\!=\!2L_0(N)$. In fact, the size of the spherical lump in the smooth phase for the $L(N)\!=\!2L_0(N)$ surface in Fig.\ref{fig-2}(b) is comparable to the half distance between the fixed boundaries, and therefore we are unable to see the surface as a string-like one. On the contrary, the lump of sphere appears sufficiently small compared with the distance  $2L(N)$ on the surfaces of $L(N)\!=\!3L_0(N)$ and $L(N)\!=\!5L_0(N)$ in Figs.\ref{fig-3}(b) and \ref{fig-4}(b), and therefore it is possible to see the surfaces as string-like when $L(N)\!=\!3L_0(N)$ and $L(N)\!=\!5L_0(N)$. 

Therefore, we expect  \cite{AMBJORN-NPB1993}
\begin{equation}
\label{tension}
Z(\alpha; L)\sim \exp(-\sigma 2L)
\end{equation}
when $L(N)\!=\!3L_0(N)$ and $L(N)\!=\!5L_0(N)$. Then, by using the scale invariance of the partition function, we have  \cite{WHEATER-JP1994,AMBJORN-NPB1993}
\begin{equation}
\label{string-tension}
\sigma = {2 S_1 - 3 N \over 2L},
\end{equation}
where $N$ and $L$ remain constant throughout the simulations. From this expression, we simply expect that $\sigma$ discontinuously changes at the transition point, because $S_1$ has a jump as we have seen in Figs.\ref{fig-7}(a)--\ref{fig-7}(c). 
 
\begin{figure}[hbt]
\centering
\includegraphics[width=12.5cm]{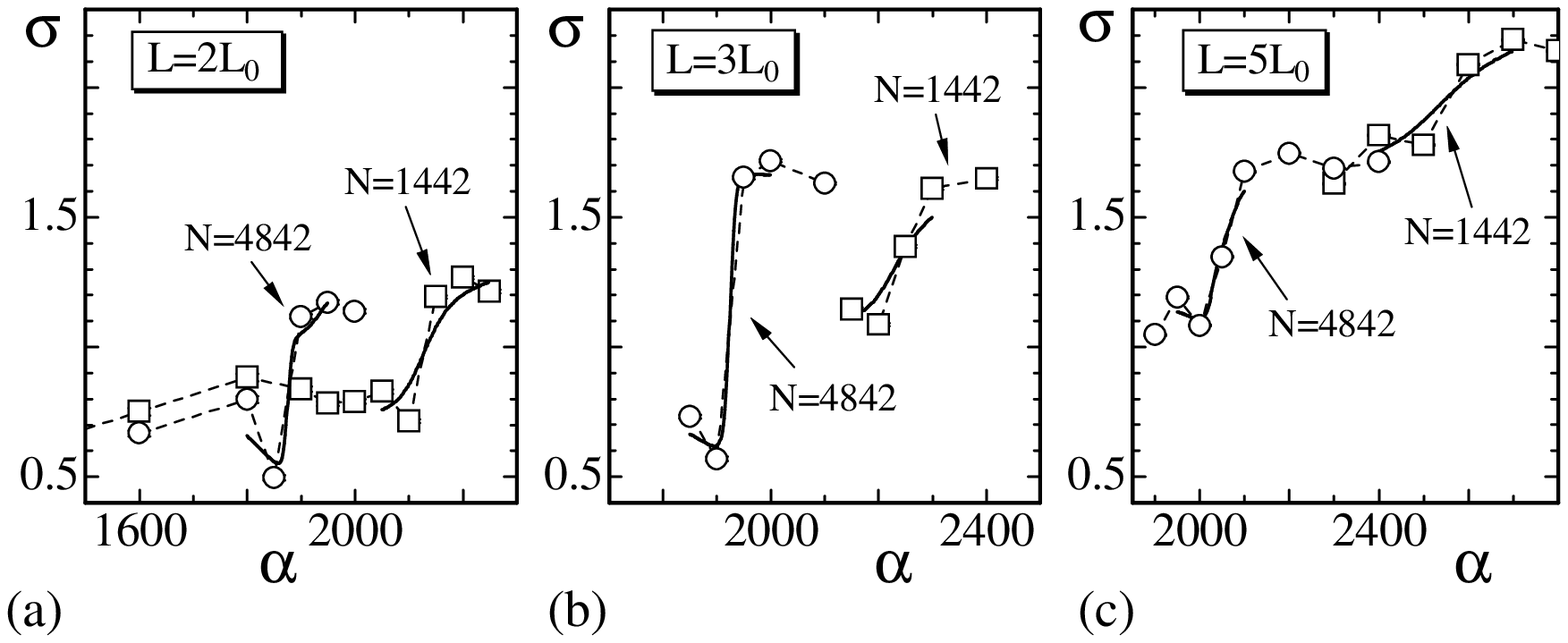}
\caption{The string tension $\sigma$ vs. $\alpha$ obtained under the conditions (a) $L(N)\!=\!2L_0(N)$, (b) $L(N)\!=\!3L_0(N)$, and (c) $L(N)\!=\!5L_0(N)$.  Solid lines were drawn by the multihistogram reweighting technique. }
\label{fig-8}
\end{figure}
Figures \ref{fig-8}(a)--\ref{fig-8}(c) are plots of $\sigma$ against $\alpha$ under the conditions $L(N)\!=\!2L_0(N)$, $L(N)\!=\!3L_0(N)$, and $L(N)\!=\!5L_0(N)$. Although $\sigma$ is not well-defined under $L(N)\!=\!2L_0(N)$ at least in the smooth phase because the surface is not always string-like as can be seen in the snapshot in Fig.\ref{fig-2}(a), it is plotted in Fig.\ref{fig-8}(a). On the contrary, $\sigma$ is considered to be well-defined in Figs.\ref{fig-8}(b) and \ref{fig-8}(c), and we see that $\sigma$ has an expected behavior against $\alpha$ under $L(N)\!=\!3L_0(N)$ and $L(N)\!=\!5L_0(N)$; $\sigma$ discontinuously changes at the first-order transition point. 

We also find from the results of the $N\!=\!4842$ surfaces shown in Figs.\ref{fig-8}(b) and \ref{fig-8}(c) that:
\begin{enumerate}
\item  $\sigma$ remains almost constant against $\alpha$ in the smooth phase
\item  $\sigma$ is independent of $L(N)$ in the smooth phase.
\item  $\sigma$ increases with increasing $L(N)$ in the crumpled phase
\end{enumerate}
On the $N\!=\!1442$ surfaces in Figs.\ref{fig-8}(b) and \ref{fig-8}(c), we find that $\sigma$ increases with increasing $N$ in the smooth phase. However, we consider this is only due to the size effect.  

\begin{figure}[hbt]
\centering
\includegraphics[width=8.5cm]{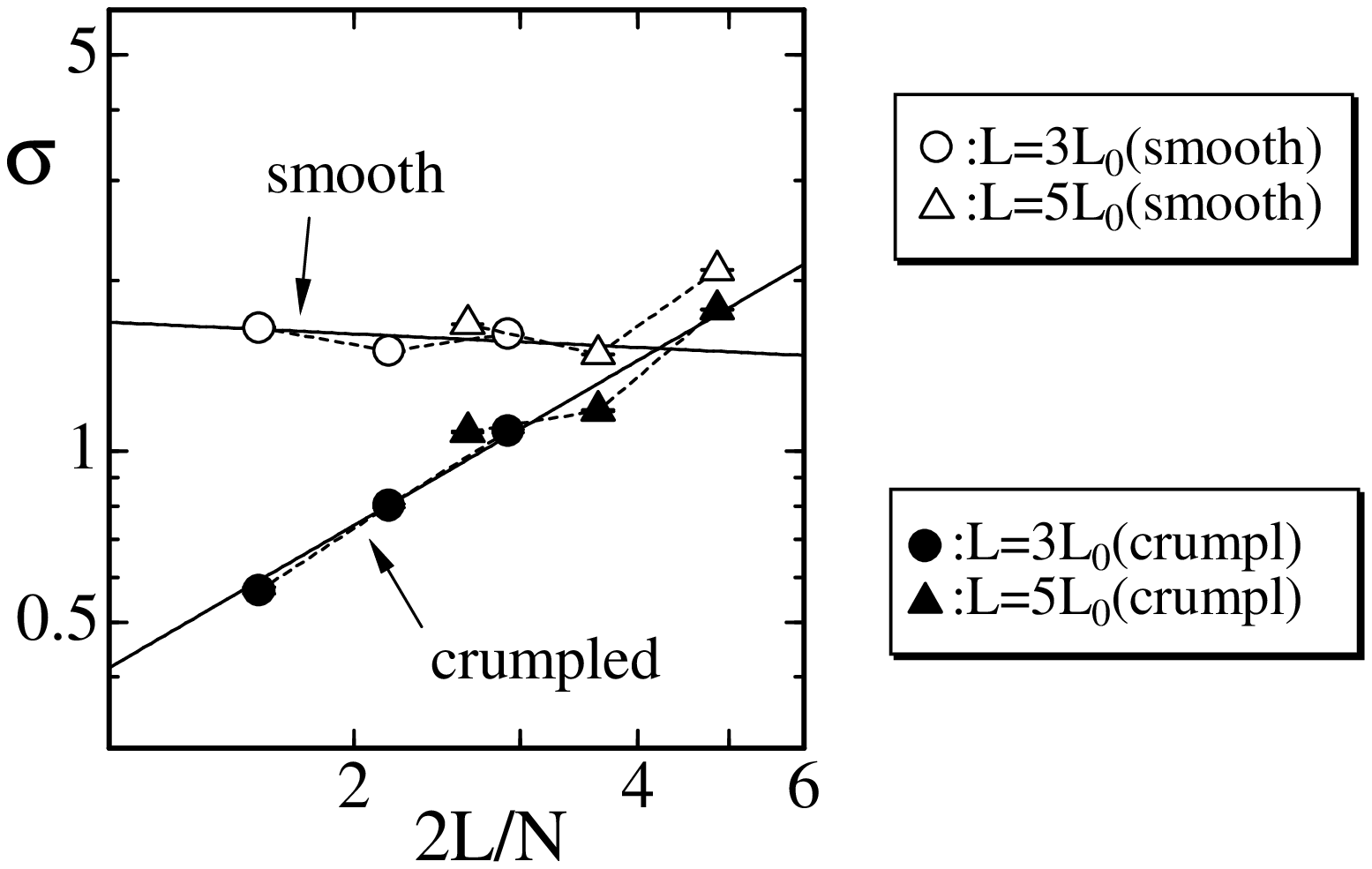}
\caption{Log-log plot of $\sigma$ against $2L/N$,  obtained at the smooth phase close to the transition point and at the crumpled phase close to the transition point. The straight lines are drawn by fitting the data to the relation of Eq.(\ref{string-tension-scaling}). }
\label{fig-9}
\end{figure}
In order to see the scaling behavior of $\sigma$ \cite{AMBJORN-NPB1993}
\begin{equation}
\label{string-tension-scaling}
\sigma \propto t^\nu,\quad t\!=\!2L/N
\end{equation}
we plot $\sigma$ versus $t\!=\!2L/N$ in Fig.\ref{fig-9} in a log-log scale, where $\nu$ in Eq.(\ref{string-tension-scaling}) is some scaling exponent. The empty (solid) symbols denote that $\sigma$ is obtained in the smooth (crumpled) phase close to the transition point of the $L\!=\!3L_0$ and  $L\!=\!5L_0$ surfaces, where $N\!=\!1442$, $N\!=\!2562$, and $N\!=\!4842$.
 The straight line drawn on the empty symbols was obtained by fitting the data to Eq.(\ref{string-tension-scaling}) excluding the largest data,  and the line indicates that $\sigma$ is constant independent of $2L/N$ in the smooth phase. Thus, we have $\nu\!\simeq\!0$ in the smooth phase. On the contrary, the exponent
\begin{equation}
\label{saling-exponent}
\nu = 0.93\pm 0.14 
\end{equation}
was obtained in the crumpled phase, and this indicates that the string tension is expected to vanish  in the crumpled phase in the limit of $t\!\to\!0$. The vanishing $\sigma$ is the property that $\sigma$ decreases with decreasing $t$ (increasing $N$ under fixed $L(N)/L_0(N)$) or  that $\sigma$ increases with increasing $L$ under fixed $N$ in the crumpled phase. Thus, the observations \newcounter{no} \setcounter{no}{1}  (\roman{no}),\setcounter{no}{2}  (\roman{no}),\setcounter{no}{3} (\roman{no}) on $\sigma$ mentioned above can be understood form the scaling property shown in Fig.\ref{fig-9}.

\begin{table}[hbt]
\caption{The curvature coefficient $\alpha$ and the length of boundary $2L$, where $\sigma$ in Fig.\ref{fig-9} were obtained. }
\label{table-1}
\begin{center}
 \begin{tabular}{cccc}
$N$  $(L/L_0)$ & $\alpha$ (crumpled) & $\alpha$ (smooth)  & $2L$ \\
 \hline
  1442 (3)    & 2200       & 2300      & 42     \\
  2562 (3)    & 2000       & 2200      & 55.8   \\
  4842 (3)    & 1900       & 1950      & 76.8   \\
 \hline
  1442 (5)    & 2500       & 2600      & 70    \\
  2562 (5)    & 2200       & 2400      & 93    \\
  4842 (5)    & 2000       & 2100      & 128   \\
 \hline
 \end{tabular} 
\end{center}
\end{table}
Table \ref{table-1} shows the curvature coefficients $\alpha$ in the crumpled phase and the smooth phase, and the boundary length $2L$, where $\sigma$ in Fig.\ref{fig-9} were obtained. $2L\!=\!42$ in the first row of the table implies that $L_0(N\!=\!1442)\!=\!7$, which is identical to that described in Section \ref{MC-technique}.

\begin{figure}[hbt]
\centering
\includegraphics[width=12.5cm]{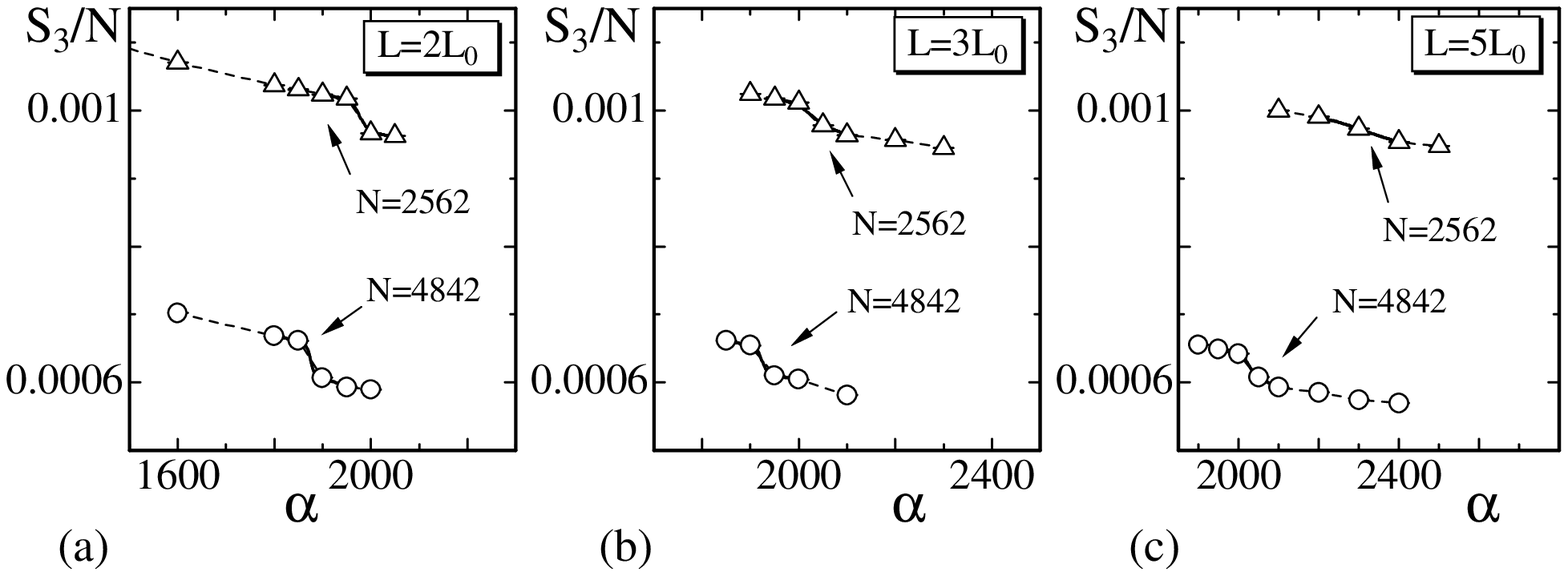}
\caption{The intrinsic curvature energy $S_3/N$ vs. $\alpha$ obtained under the conditions (a) $L(N)\!=\!2L_0(N)$, (b) $L(N)\!=\!3L_0(N)$, and (c) $L(N)\!=\!5L_0(N)$. Solid lines were drawn by the multihistogram reweighting technique.  }
\label{fig-10}
\end{figure}
The first-order transition should be reflected in $S_3/N$ such that $S_3/N$ discontinuously changes at the transition point. In order to see this, we plot $S_3/N$ in Figs.\ref{fig-10}(a)--\ref{fig-10}(c). We find that $S_3$ has a jump or a gap under the conditions $L(N)\!=\!2L_0(N)$, $L(N)\!=\!3L_0(N)$, and $L(N)\!=\!5L_0(N)$. The discontinuity of $S_3/N$ in the $N\!=\!4842$ surface is more clear than those in the smaller surfaces, and this also shows that the model undergoes a first-order transition. 

We should note that a possibility of continuous transition is not completely eliminated, although $S_1/N$, $\sigma$, and $S_2/N_B$ appear to change discontinuously in our simulation data. In order to see this more clearly, we need more simulation data at the transition point. Then, we can find the order of the transition precisely by using the finite-size scaling analysis. 
 
\section{Summary and conclusions}
We have investigated an intrinsic curvature model on dynamically triangulated spherical surfaces by fixing two vertices as point boundaries separated by the distance $2L(N)$. Three different values are assumed for $L(N)$;  $L(N)\!=\!2L_0(N)$, $L(N)\!=\!3L_0(N)$, and $L(N)\!=\!5L_0(N)$, where $L_0(N)$ is the radius of a sphere of size $N$ and is chosen such that $S_1/N$ is almost identical to $S_1/N\!=\!1.5$. The surface size are $N\!=\!1442$, $N\!=\!2562$, and $N\!=\!4842$. 

A deficit angle term $S_3$ is assumed for the curvature energy, and the Gaussian bond potential $S_1$ is included in the Hamiltonian; $S\!=\!S_1\!+\!\alpha S_3$, where $\alpha$ is the curvature coefficient. The model is expected to have the smooth phase at sufficiently large $\alpha$ on dynamically triangulated surfaces. It is also expected that the smooth phase is separated from a non-smooth phase by the first-order transition, where the surface is almost linear and is very similar to the branched polymer surface \cite{KOIB-PRE-2003}. We also confirmed in \cite{KOIB-JSTAT-2006-1} that the order of the transition changes from first-order to second-order on tethered surfaces as the distance $L(N)$ increases. Therefore, we aimed in this paper to see how the transition changes depending on $L(N)$ on dynamically triangulated fluid surfaces, where the surfaces are expected to be oblong at sufficiently large $L(N)$. 

It was confirmed in this paper that the first-order transition remains unchanged and is not influenced by how large $L(N)$ is, up to $L(N)\!=\!5L_0(N)$ at least. In the crumpled phase, the surface spans from one boundary point to the other as an oblong and almost linear surface under the conditions $2L_0(N)\!\leq\! L(N)\!\leq\!5L_0(N)$. On the other hand, there appears a spherical lump on the surface between the two boundaries in the smooth phase. Almost all vertices are included in the lump, and therefore physical quantities, such as the bending energy $S_2/N_B$, change depending on whether the lump is included in the surface or not. A discontinuous change of $S_2$, which is not included in the Hamiltonian, characterizes the first-order transition. The deficit angle term in the Hamiltonian also changes discontinuously at the transition point. 

The string tension $\sigma$ was obtained when $L(N)\!=\!3L_0(N)$ and $L(N)\!=\!5L_0(N)$, where the surface is sufficiently oblong even in the smooth phase though the lump is included in the surface.  We found that $\sigma$ has a jump at the transition point, and that $\sigma$ takes a constant value in the smooth phase in the sense that it is independent of the distance $L(N)$. On the contrary, $\sigma$ increases with increasing $L(N)$ in the crumpled phase.

The scaling property $\sigma\!\propto\!(2L/N)^\nu$ was also obtained at the smooth phase and at the crumpled phase close to the transition point. We have $\nu\!\simeq\! 0$ in the smooth phase, and $\nu \!=\! 0.93\pm 0.14$ in the crumpled phase. Thus, the crumpled phase is characterized by vanishing string tension; $\sigma$ decreases with decreasing $t\!=\!2L/N$  (increasing $N$ under fixed $L(N)/L_0(N)$) or equivalently $\sigma$ increases with increasing $L$ under fixed $N$ in the crumpled phase. This describes the above mentioned dependence of $\sigma$ on $L(N)$ in the crumpled phase.

It is interesting to calculate $\sigma$ in the extrinsic curvature model on relatively large fluid surfaces. The shape of surfaces of the extrinsic curvature model \cite{KOIB-PLA-2004} seems very different from those of the intrinsic curvature model, and the dependence of the transition on $L$ is also different from each other; the transition strengthens (remains unchanged) as $L$ increases in the extrinsic (intrinsic) curvature model. Moreover, the dependence of $\sigma$ on the curvature coefficient in both models is almost identical with each other; $\sigma$ has a jump at the transition point under sufficiently large $L$ in both models.

\section*{Acknowledgment}
This work is supported in part by Grant-in-Aid of Scientific Research, No. 15560160 and No. 18560185. The author (H.K.) is thankful to a referee for helpful comments and suggestions.



\end{document}